# Dynamic Contraction of the Positive Column of a Self-Sustained Glow Discharge in Air Flow

M.N.Shneider
*Princeton University, Princeton, NJ, 08544*

M.S.Mokrov
*Institute for Problems in Mechanics, RAS, Moscow*, Russia

and

G.M.Milikh
*University of Maryland, College Park, MD, 20742*

**We study the dynamic contraction a self-sustained glow discharge in air in a rectangular duct with convective cooling. A two dimensional numerical model of the plasma contraction was developed in a cylindrical frame. The process is described by a set of time-dependent continuity equations for the electrons, positive and negative ions; gas and vibrational temperature; and equations which account for the convective heat and plasma losses by the transverse flux. Transition from the uniform to contracted state was analyzed. It was shown that such transition experiences a hysteresis, and that the critical current of the transition increases when the gas density drops. Possible coexistence of the contracted and uniform state of the plasma in the discharge, where the current flows along the density gradient of the background gas, is discussed.**

Contraction (also called constriction) of a gas discharge occurs when current contracts from a significant volume of weakly ionized plasma into a thin channel and has attracted attention of scientists since the late sixties. Studies of the contraction mechanisms, besides being interesting science, are of practical importance, especially for applications when a uniform gas flow is desirable. Usually, the contraction has been studied in a tube with diffusive cooling, or in a weakly ionized gas flow in a plane duct. Analysis and modeling of contraction which occurs in the tube were first conducted in [1,2]. The qualitative features of such process were obtained for different gases [3,4], and the fine effects taking place during the transition from the diffusive to constricted phase were studied in [4-8]. Presently, 1D models of current contraction in a tube are well developed. They take into account the kinetics of electrons, ions, and metastable particles, and provide reasonable agreement with the experiments [4-8].

There are a number of experiments stimulated by the development of powerful gas-discharge lasers and plasma-chemical reactors (see review [9]) where the current contraction phenomenon was observed when the discharge power exceeds a certain threshold level. However, this phenomenon in a plane duct in the presence of the gas flow is not well understood. Two questions usually emerge: what is the critical current (or current density), i.e maximum current that allows uniform discharge to be stable, and what are the dynamics of the transition from the uniform to contracted phase. The critical current (or critical power input) was found in the earlier papers (see [9]). The instability is caused by the thermal-ionization mechanism [10] that is typical for molecular gases including nitrogen and air.

Previous research has examined the build-up dynamics of the contracted state. The experiments show that the formation of the contracted state occurs nonuniformly [9, 11−18]. This process involves growing the current filament from one electrode to another, and like a streamer, the filament eventually cross-connects the electrodes of discharge gap. This impressive phenomenon was extensively studied in [9, 11-22]. A number of questions arise such as: is the growing filament analogous to the ionization wave type of a "thermal" streamer? What are the plasma properties behind the front of the rising channel? Does the field strengthening and increasing current which occur at the filament tip play an important role for the channel expansion? Those questions were discussed in the previously mentioned papers, where analytical and numerical calculations were conducted mostly for non-self-sustained discharge. The main conclusions were that: a thermal streamer does not exist, the strengthening of the field and increase of the tip current are not essential, and the state of the plasma behind the front is determined mainly by the energy input. However, up to now, for self-sustained discharge, there is no presently comprehensive understanding of the dynamics of non-uniform contraction including formation of the steady state. In addition, the questions of why the nonuniform contraction occurs, and what are the required conditions, have never been asked.

Our earlier work [23,24] that conducted 2D numerical modeling of the dynamic contraction in molecular nitrogen in a plane geometry, was an attempt to address some of these issues. The self-sustained gas discharge was treated in a quasineutral approximation and the contraction was due to instability in the positive column of the discharge. It was shown that this phenomenon is related to the fact that transition of the uniform state to the contracted state allows hysteresis, i.e. both of the states coexist in the same current. By applying a strong enough perturbation, one abruptly transforms the uniform state into contracted. The transformation begins where the perturbation was created and eventually fills the whole gap. There is a of 2D "contraction wave" front that moves in space, while the plasma density, along with the translational and vibrational temperature increases behind the wave front. The objective of this work is to model the observed motion of the contraction front in air and to determine the conditions required for this effect to exist.

Analysis of the current contraction in the gas flow is laborious since the discharge channel is always 3D due to the presence of the gas flow. Even 2D model of the process is quite complex, in the case of 3D model the complexity increases and become cumbersome. Thus in this paper the 3D problem will be reduced to 2D cylindrical model or 1D cylindrical model, where the longitudinal coordinate along the current is excluded from consideration. Furthermore, in order to take into account the plasma and energy losses in the contracted channel due to the gas flow, we introduce an effective rate of particle and energy removal. In another assumption, we ignore the dynamics of near-electrode layers. The role played by the layers in our analysis is limited to be a source of perturbations for the whole plasma volume. The shape and character of such perturbations deserve special treatment that is beyond the scope of the present work.

The study presented in this paper was stimulated by the fact that there are a large number of experiments on the dynamic contraction of a glow discharge in nitrogen and air. The peculiarity of the air discharge is associated with the presence of negative ions and related processes of electron attachment to molecular oxygen and electron detachment from the negative ions. Similar processes play a role in the powerful gas-discharge lasers since the latter are filled mostly with electronegative gases.

**I. Basic model for spatial-temporal dynamics of contraction in the flow of weakly ionized air**

Recently we presented a theoretical model for the dynamic contraction of the current channel inside a quasineutral positive column of a self-sustained glow discharge in nitrogen. The process occurs in a rectangular duct with convective cooling [23,24] and is described by the 2D model of current contraction. If the contraction develops in a tube, [2,5,8] a cylindrical model with axial symmetry is more accurate. For experiments [9,16] where the convective cooling was used, the contracted channel has a more complex shape and, as mentioned above, a full 3D model is more relevant. This is why in [24] we analyzed the contraction development in nitrogen flow in 2D cylindrical model with the axial symmetry, and then compared it with the 2D plane model [23,24]. We found that the models are in agreement with each other and with experiments [16].

As in [23-25] let us consider a column of a convection-cooled glow air discharge that occurs in a rectangular duct loaded by the DC circuit with a ballast resistance (Fig. 1,a). We will describe the process of contraction development in the axis-symmetrical cylindrical geometry (Fig. 1,b). Here we will take into account the convective thermal and plasma losses by introducing effective convective lifetime of the gas in the discharge gap. We recall that in reality the contracting channel is shifted sideways by the gas flow having the velocity **u**. (Fig. 1,a), which shows three consecutive snapshots of the contracted plasma discharge, and thus a 3D description of the process could be more adequate to the task of simulations of the dynamic contraction.

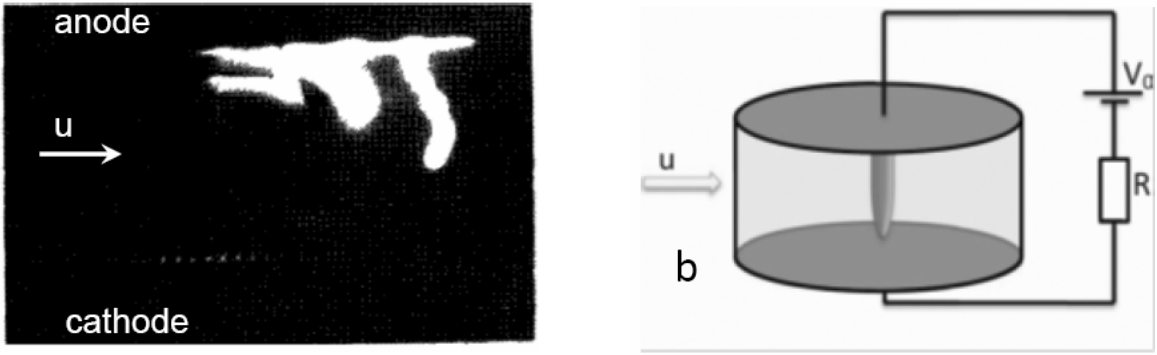

**Fig.1**. (a) Snapshots of the plasma discharge with contraction that occurs in weakly ionized air flow [9]. (b) Contracting channel in axisymmetric cylindrical geometry (x,r) in our model. The arrow shows the gas flow with the velocity u.

The discharge model will be treated by assuming quasineutrality of the plasma. The respective set of equations consists of continuity equations for the densities of electrons, $n_e$, positive $n_+$ and negative $n_-$ ions, equation for the electric field **E**, gas translational, $T$, and vibrational, $T_V$, temperatures and equation for the electric circuit. The continuity equations for $n_e$, $n_+$ and $n_-$ are:

$$\frac{\partial n_e}{\partial t} + \nabla \cdot \mathbf{\Gamma}_e = Q_e, \tag{1}$$

$$\frac{\partial n_+}{\partial t} + \nabla \cdot \mathbf{\Gamma}_+ = Q_+, \tag{2}$$

$$\frac{\partial n_-}{\partial t} + \nabla \cdot \mathbf{\Gamma}_- = Q_-, \tag{3}$$

where $\mathbf{\Gamma}_e = -n_e \mu_e \mathbf{E} - D_e \nabla n_e$, $\mathbf{\Gamma}_+ = n_+ \mu_+ \mathbf{E}$, and $\mathbf{\Gamma}_- = -n_- \mu_- \mathbf{E}$ are the corresponding fluxes of charged particles; $\mu_e, \mu_+, \mu_-$ are their mobilities; $D_e$ is the electron diffusion coefficient; $\tau$ is the characteristic time for convective removal of charged particles (and heat) from the discharge. Source terms in right-hand sides of (1) - (3) are: $Q_e = (\nu_{ion} - \nu_a)n_e + \nu_d n_- - \beta_{e+} n_e n_+ - \frac{n_e}{\tau}$, $Q_- = \nu_a n_e - \nu_d n_- - \beta_{ii} n_- n_+ - \frac{n_-}{\tau}$, $Q_+ = Q_e + Q_-$.

Here $\nu_{ion}, \nu_a, \nu_d$ are the ionization, attachment and detachment frequencies; $\beta_{e+}, \beta_{ii}$ are the coefficients for electron-ion and ion-ion recombination; in the plane geometry $\tau \approx L_z/u$ is the characteristic time of convective plasma cooling with laminar gas flow parallel to the electrodes ($L_z$ is the characteristic scale length along the flow), and $u$ is the flow velocity. In the assumed cylindrical geometry $\tau \approx 2r_{max}/u$, where $r_{max}$ is the radius of the domain occupied by the weakly ionized plasma. Here we neglect a small displacement of the contracted channel due to the convective gas flow.

If the quasi-neutrality condition $n_+ = n_e + n_-$ is fulfilled one of the equations (1)-(3) becomes redundant. Therefore we have solved Eqs. (1) and (3) for $n_e$ and $n_-$, while $n_+$ was obtained as a sum of $n_e + n_-$ in each point.

The electric field $\mathbf{E} = -\nabla \varphi$ is determined from the charge conservation equation

$$\nabla \cdot \mathbf{j} = 0, \quad \mathbf{j} = -e(\mathbf{\Gamma}_e - \mathbf{\Gamma}_+ + \mathbf{\Gamma}_-). \tag{4}$$

A significant fraction of the power released in a discharge in air is pumped to the vibrational degrees of freedom. However, in view of rapid VT relaxation of oxygen molecules, we assumed that in the air the vibrational energy is stored in the nitrogen molecules. Thus the balance equation for the vibrational energy, $E_V$ yields:

$$\frac{\partial E_V}{\partial t} - \frac{\partial}{\partial x}\left(D\frac{\partial E_V}{\partial x}\right) - \frac{1}{r}\frac{\partial}{\partial r}\left(r\left(D\frac{\partial E_V}{\partial r}\right)\right) = \eta_V \mathbf{j}_e \mathbf{E} - \hbar\omega\Pi - \frac{E_V - E_V^0}{\tau_{VT}} - \frac{(N_0/N)[E_V - E_V^0(T_0)]}{\tau} \quad (5)$$

Here $D$ is the diffusion coefficient for $N_2$ molecules, $E_V = x_{N2} N \hbar\omega_{N2} / (\exp(\hbar\omega_{N2}/kT_V) - 1)$, $\hbar\omega_{N_2} = 0.292$ eV is the vibrational quanta of $N_2$ molecules; $E_V^0$ is the equilibrium value of $E_V$; $\Pi$ is the flow of vibrational quanta in the vibrational energy domain, which describes the energy sink that brings the upper vibrational levels down; $\eta_v$ is the fraction of Joule heat responsible for the vibrational excitation; $\tau_{VT}$ is the VT relaxation time of the vibrational levels; $x_{N2} = 0.8$ is the $N_2$ fraction, $T_0$=293 K, $N_0 = p/(kT_0)$, $p$ is the gas pressure. The expression for the flow $\Pi(T,T_v)$ in Eq. (5) is adapted from [26].

As the pressure is equalized quickly, we can assume that the gas is heated under isobaric conditions. In this case the gas density $N$ and temperature $T$ are related as $N \propto T^{-1}$. For $T$ we get the equation:

$$Nc_{p1}\frac{\partial T}{\partial t} - \frac{\partial}{\partial x}\left(\lambda(T)\frac{\partial T}{\partial x}\right) - \frac{1}{r}\frac{\partial}{\partial r}\left(r\left(\lambda(T)\frac{\partial T}{\partial r}\right)\right) = \eta_T \mathbf{j}_e \mathbf{E} + \hbar\omega\Pi + \frac{E_V - E_V^0}{\tau_{VT}} - N_0 c_{p1}\frac{T - T_0}{\tau} \quad (6)$$

Here $c_{p1}$ is the specific heat of air at a constant pressure; $\lambda(T)$ is the molecular heat conductivity; $\eta_T = 1 - \eta_V$ is the fraction of Joule heat responsible for the direct heating of the gas.

The set of equations (1), (3) - (6) is solved jointly with the time-dependent equation for the electric circuit:

$$\frac{dQ}{dt} = \frac{V_0 - V}{R} - I \quad (7)$$

where $V_0$ is the power supply voltage; $R$ is the external ballast resistor; $I$ is the discharge current; $Q = \varepsilon_0 \int E dS$ is the total charge of a metallic electrode, $\varepsilon_0$ is the dielectric permittivity of vacuum; $E$ is the electric field near the electrode.

The calculations were performed for air at $p$=100 Torr and for the following set of parameters. The electron and ion mobility: $\mu_e p = 0.45 \times 10^6$ cm$^2$ Torr V$^{-1}$ s$^{-1}$, while

$$\mu_+ = \mu_- = \begin{cases} \dfrac{2100}{p[\text{Torr}]}, & E/p < 49.3 \quad \text{V/(cm Torr)} \\ \dfrac{14745}{p\sqrt{E/p}}, & E/p > 49.3 \quad \text{V/(cm Torr)} \end{cases}$$

The diffusion coefficient for electrons, $D_e = \mu_e T_e$, where $T_e = T_e(E/N) = \frac{2}{3}\varepsilon(E/N)$, and $\varepsilon(E/N)$ is the average electron energy adapted from [27]. The plasma kinetic parameters of the air are taken similar to [28- 30]. The ionization frequency: $\nu_i = (K_{N2} x_{N2} + K_{O2}(1 - x_{N2}))N$,

$K_{N_2}[\text{cm}^3/\text{s}] = 8.1 \times 10^{-9} \exp(-925/(E/N[Td]))F$,

$K_{O_2}[\text{cm}^3/\text{s}] = 4.9 \times 10^{-9} \exp(-657/(E/N[Td]))$

where the coefficient $F$ takes into account the effect on the ionization rate by the vibrationally excited nitrogen molecules, $F = \exp(Cz/(E/N)^2)$, $C = 6.5 \times 10^3$ Td$^2$, $z = \exp(-\hbar\omega_{N_2}/kT_V)$ [31]. Since in all the cases discussed the value of electron temperature does not exceed 0.8 – 1.5 eV for the entire range of $E/N$, we consider only the electron attachment in the three-body collisions: $e + O_2 + M \rightarrow O_2^- + M$, $M = O_2, N_2$ with the effective attachment frequency [28,29] $\nu_a = (k_{aO_2}(1-x_{N2})^2 + k_{aN_2}(1-x_{N2})x_{N2})N^2$,

$k_{aO_2}$ [cm$^6$s$^{-1}$] = $1.4 \times 10^{-29}(300/T_e)\exp(-600/T)\exp(700(T_e-T)/(T_eT))$

$k_{aN_2}$ [cm$^6$s$^{-1}$] = $1.07 \times 10^{-31}(300/T_e)^2\exp(-70/T) \exp(1500(T_e-T)/(T_eT))$,

where $T_e$ and $T$ are in Kelvins. The detachment occurs in ion collisions with neutrals, it has the effective frequency $\nu_d = (k_{d,O2}(1-x_{N2}) + k_{d,N2}x_{N2})N$ [1/s], where

$k_{d,O_2}[cm^3 s^{-1}] = 2.7 \times 10^{-10}(T/300)^{0.5}\exp(-5590/T)$,

$k_{d,N_2}[cm^3 s^{-1}] = 1.9 \cdot 10^{-12}(T/300)^{0.5}\exp(-4990/T)$. The coefficients for electron-ion and ion-ion recombination are:

$$\beta_{e+}[cm^3 s^{-1}] = 2 \cdot 10^{-7}(300/T_e)^{0.5},$$

$$\beta_{ii}[cm^3 s^{-1}] = 2 \cdot 10^{-7}(300/T)^{0.5}(1+10^{-18}N(300/T)^2),$$

where $T_e$ and $T$ are in Kelvins and $N$ is given in cm$^{-3}$. The vibrational relaxation time:

$\tau_{VT}[s] = N^{-1}(7 \cdot 10^{-10} \exp(-141/T^{1/3}) + 5 \cdot 10^{-12} x_0 \exp(-128/T^{1/2}))^{-1}$, where $N$ is in cm$^{-3}$ and $x_0$ is the fraction of O atoms. It was assumed in all the computed cases that $x_0 = 10^{-4}$. Heat capacity at constant pressure, $c_{p1} = 7/2k$. It was also assumed that $\eta_V = 0.95$, which is in agreement with the simulations [32,10] made for the air and for the range of reduced electric field $E/N$ which are typical for the conditions of our model.

Similar to the model of contraction in nitrogen [23,24] we assume that

$$p = NkT = \text{const}, \tag{8}$$

where $k$ is the Boltzmann constant.

We used the following boundary conditions for the equation set (1), (3) - (6): since the near-electrode region is excluded from consideration, we assume that $\partial n_e/\partial x = \partial n_-/\partial x = \partial T/\partial x = \partial E_V/\partial x = 0$ on the electrodes boundaries (at $x = 0$, $x=L_x$). For the electric potential we have $\varphi = 0$ at $x = 0$ and $\varphi = V$ at $x=L_x$, where $V$ is the discharge voltage. It is assumed that $\partial n_e/\partial r = \partial n_-/\partial r = \partial T/\partial r = \partial E_V/\partial r = \partial \varphi/\partial r = 0$ on the symmetry axis ($r = 0$) and on the side wall of the discharge chamber ($r = r_{max}$).

**III. Transient Conditions**

It is shown in [23- 25] that transition to the contracted state in nitrogen occurs in hysteresis mode, often called "hard mode". Note that the hysteresis type transition from diffusive to contracted state that which occurs in a tube was recently studied in [5,7,26], where the diffusive and contracted states coexisted along the current direction. Now we consider the transition from uniform to contracted glow discharge in the weakly ionized air plasma flow for the conditions similar to experiments [16].

The current-voltage characteristic of the glow discharge of the positive column are found from the condition

$$E = I / \int_0^{r_{max}} 2\pi r\sigma dr, \tag{9}$$

where $\sigma = e[\mu_e n_e + \mu_+(n_e + n_-) + \mu_- n_-]$ is the conductivity of the quasi-neutral air plasma;. Simultaneously the one-dimensional radial dependent plasma density balance equations (1)-(3) and equations (5) for $E_V$ and (6) for $T$ are solved, assuming the discharge uniformity along the $x$ axis. The radius of the cylindrical chamber is taken as $r_{max}$= 2 cm and the characteristic time for convective heat removal is assumed to be $\tau = 10^{-3}$ s. For a given current $I$, the equation set was solved by the relaxation method. The calculations were performed for arbitrary homogeneous initial distributions of $T_V, T$ and plasma components.

A set of 1D radial dependent equations for a weakly ionized air plasma contained in a cylinder chamber with the radius $r_{max}$= 2 cm and with the transverse gas flow has been solved under the assumption that the characteristic time for convective heat removal $\tau = 10^{-3}$ s. "Current-voltage characteristics" $E(I)$ obtained for different gas pressures (densities) are shown in Fig. 2. The top (blue) traces correspond to the uniform state of the air plasma while the bottom (red) traces correspond to the contracted states. It is apparent that hysteresis transition becomes more pronounced when the pressure drops. The computed values of the critical discharge current $I_{cr}$ required for the transition from the uniform to contracted state increase when the pressure drops. This is consistent with the result found in the plane geometry [23-25].

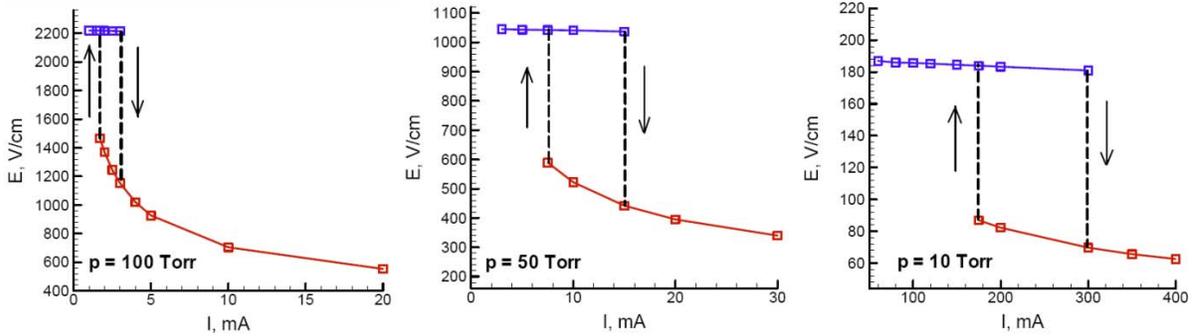

**Fig. 2.** The "current-voltage characteristic" of the glow discharge in air flow at the different pressures. The critical current $I_{cr}$ is shown by the bold trace. If $I < I_{cr}$ the discharge is uniform, if $I > I_{cr}$ the contracted channel is formed.

Figure 3 shows two current-voltage characteristics of the glow discharge both obtained for 50 Torr air pressure, but under different convective heat removal time (squares correspond to $\tau = 10^{-3} s$, triangles correspond to $\tau = 5 \times 10^{-3} s$). The figure reveals that for the fast convective heat removal ($\tau = 10^{-3} s$) the hysteresis transition occurs at the current 1.6 times higher than that for the slow convective heat removal ($\tau = 5 \times 10^{-3} s$).

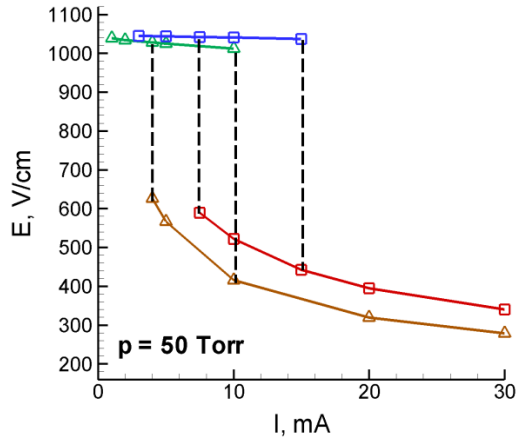

Fig. 3. Comparison of the two current-voltage characteristics of the glow discharge obtained under different convective heat removal time (squares correspond to $\tau = 10^{-3} s$, triangles correspond to $\tau = 5 \times 10^{-3} s$).

The results shown in this Section imply that if the gas density is nonuniform and the current flows along the density gradient the contracted and diffusive states could exist in different regions of the discharge. This regime occurs when the current exceeds its critical amplitude in the high density region, while remaining below the critical amplitude in the low density region. This is similar to the existence of diffusive and contracted states in the long tube [5,8] (at constant gas density). In other words, at a given current $I$, if the contracted channel grows from the high pressure region where $I > I_{cr}(N)$ to the low pressure region, the wave of the thermal-ionization instability stops as soon as the air density drops to the value $\widetilde{N}$ such that $I = I_{cr}(\widetilde{N})$.

Notice that the dependence of the critical current on the gas pressure is due to the stabilizing role played by the ambipolar diffusion and by V-T relaxation, which is reduced under the low pressure.

### IV. 2D axes symmetric computations for the dynamics of contraction occurred in cylindrical geometry.

Let us assume that the discharge gap is a cylindrical chamber having the length 2 cm, and the radius $r_{max} = 2$ cm. Furthermore assumethe gas pressure p=50 Torr. Under this conditions and at a current of 13.8 mA the simulations give an uniform initial quasi-neutral plasma with the density $7.65 \cdot 10^9$ cm$^{-3}$ (the electron density is $0.65 \cdot 10^9$ cm$^{-3}$ while density of the negative ions is $7 \cdot 10^9$ cm$^{-3}$); the gas temperature is slightly higher than the room temperature; and the vibrational temperature is $T_v = 980$ K. Here the computations were made in the point of the volt ampere characteristic of the 1D discharge which is stable with respect to small perturbations (see Fig.2). In order to generate contraction one has to apply a perturbation of sufficient amplitude. Thus we perturb the uniform initial state by applying the following artificial temperature distributions:

$$T(x,r) = 293 \cdot (1 + 1 \cdot \exp(-r^2/0.15^2) \cdot \exp(-(x-d)^2/0.2^2)) \text{ K},$$
$$T_v(x,r) = 1069.5 \cdot (1 + 1 \cdot \exp(-r^2/0.15^2) \cdot \exp(-(x-d)^2/0.2^2)) \text{ K} \quad (10)$$

and will integrate the equation set (1)-(8) taking (10) as the initial conditions. Furthermore, we will load the discharge gap by external circuit with $V_0 = 9$ kV, $R = 500$ kOhm.

The following set of panels 4a-4d shows 2D temporal evolution of the discharge that came as result of our numerical simulations.

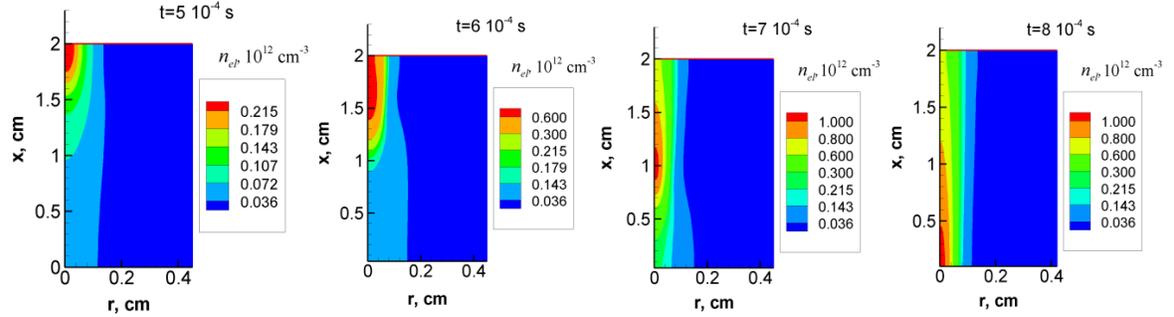

**Fig. 4a**. Electron density distributions at different times

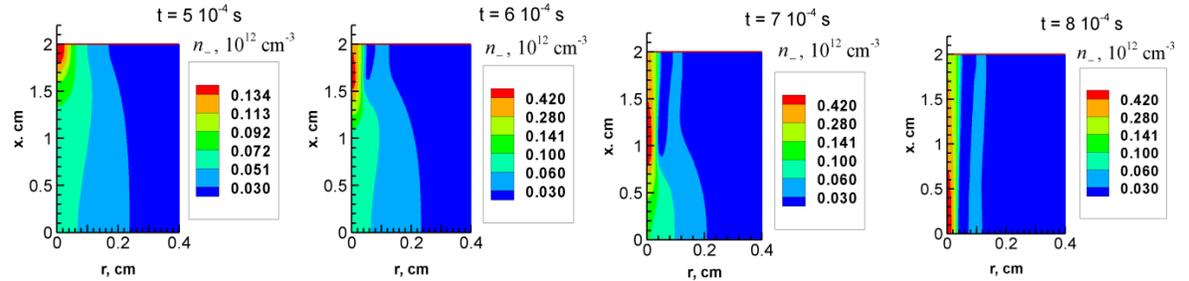

**Fig. 4b**. Negative ion density distribution at different times

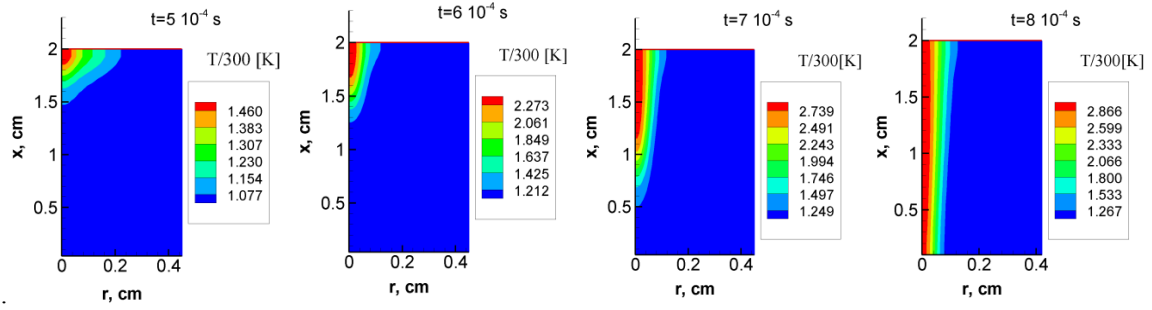

**Fig. 4c**. Translational temperature at different times

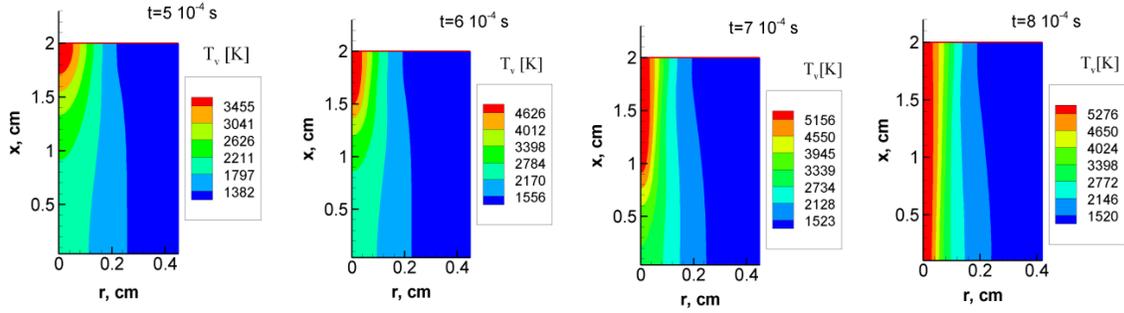

**Fig. 4d**. Vibrational temperature at different times

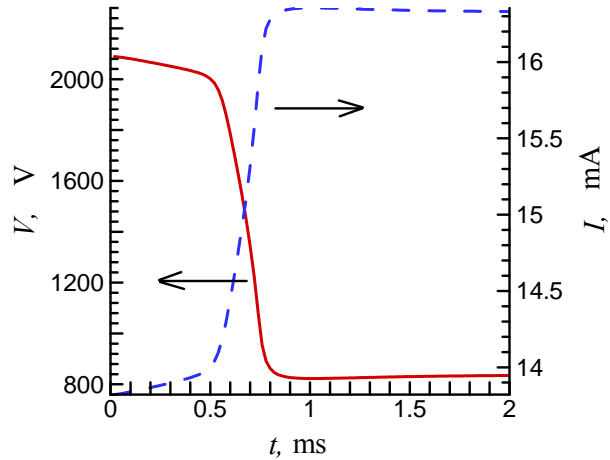

**Fig.5**. The transition evolution of the plasma column voltage, and the discharge current in the hysteresis region for the case shown in Fig. 4.

Figure 5 shows the computed temporal evolution of the plasma column voltage and the discharge current in a process of transition from the uniform to the contracted regime for the case shown in Fig. 4. Here the steady discharge current is 16.3 mA and the electric field in the column discharge is 418 V/cm which is in agreement with the earlier 1D cylindrical model.

The corresponding 1D distribution of plasma parameters along the axes of symmetry ($r=0$) a different times is shown in Fig. 6.

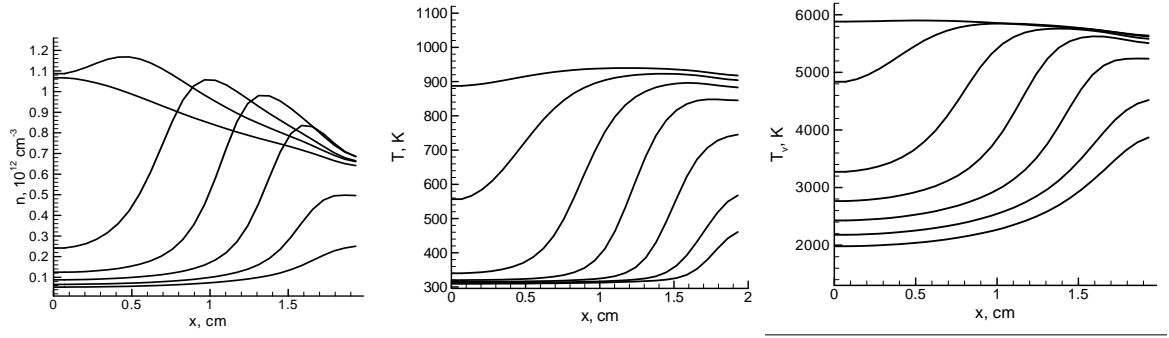

**Fig. 6.** Distribution of the plasma parameters along the current on the symmetry axis ($r=0$): the electron density (a), the translational (b) and the vibrational (c) temperatures in time interval between 0.5 and 0.8 ms (from the bottom to the top) with the time step 0.05 ms. Initial perturbation was set near $x = 2$ cm.

Our simulations show that at the discussed conditions the velocity of the contraction channel grows in time from tens to hundreds m/s. This agrees with the known experiments (see for example [16]).

Radial distribution of the charge density (in $10^{12}$ cm$^{-3}$) in the middle of the discharge gap in the steady regime, $t \geq 8 \cdot 10^{-4}$ s, are shown in Fig. 7:

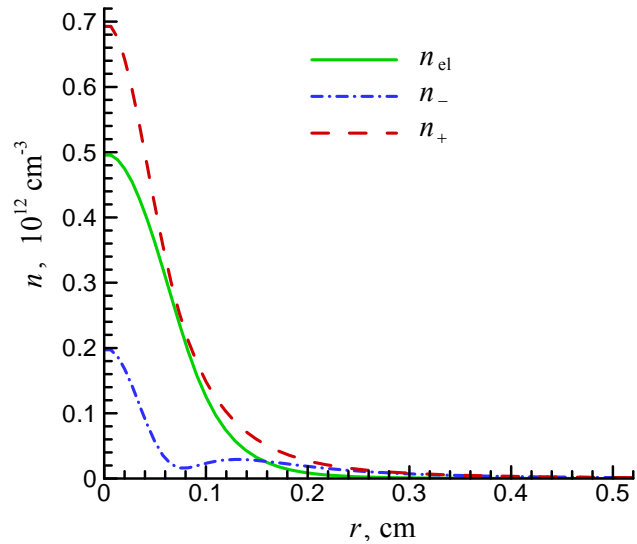

**Fig. 7.** The radial distributions of the plasma charge densities at $t \geq 8 \cdot 10^{-4}$ s.

## Conclusions

For the first time, a self-consistent 2D model of the current contraction in air has been developed. The model takes into account a number of relevant processes including formation and decay of negative ions, vibrational non-equilibrium and external electric circuit, and describes a relatively slow mechanism of the thermal-ionization instability in a weakly ionized plasma flow.

It was shown that transition from the diffuse to the contracted state occurs in hysteresis mode. The critical current for the transition depends on the gas pressure due to the stabilized role played by the ambipolar diffusion, which increases when the pressure drops, and by the V-T relaxation, which reduces with the pressure.

The critical current increases when the pressure (density) of the background gas reduces. Therefore, in the discharge that occurs in a nonuniform gas, the contracted and diffuse states could coexist if the current is flowing along the density gradient.

The discussed problem has implications to the studies of powerful gas-discharge lasers and of plasma-assisted combustion.

**Acknowledgements**

This work was supported in part by AFOSR MURI grant FA9550-09-1-0602 with Ohio State University overseen by Dr. Chiping Li.**References**

[1] G. Ecker, W. Kroll, K. H. Spatschek, and O. Zoller, "Negative characteristic and instability of collision-dominated helium plasma column" *Phys. Fluids* **10**, 1037 (1967).
[2] E.F. Jaeger, L. Oster, and A.V. Phelps, "Growth of thermal constrictions in a weakly ionized-gas discharge in helium", *Phys. Fluids* **19**, 819 (1976)
[3] A.V. Eletskii, B.M. Smirnov, "Nonuniform gas discharge plasma" *Phys.Usp.* **39** 1137–1156 (1996)
[4] Yu B Golubovskii, V Nekuchaev, S Gorchakov and D Uhrlandt, "Contraction of the positive column of discharges in noble gases" *Plasma Sources Sci. Technol.* **20** 053002 (2011).
[5] Yu.Z.Ionikh, A.V. Meshchanov, F. B. Petrov, N. A. Dyatko, and A. P. Napartovich "Partially constricted glow discharge in an argon–nitrogen mixture", *Plasma Physics Reports*, **34,** N 10, pp. 867–878 (2008).
[6] N. A. Dyatko, Y. Z. Ionikh, I. V. Kochetov, D. L. Marinov, A.V. Meshchanov, A.P. Napartovich, F.B. Petrov and S.A. Starostin "Experimental and theoretical study of the transition between diffuse and contracted forms of the glow discharge in argon", *J. Phys. D: Appl. Phys.* **41** 055204 (2008).
[7] I. A. Shkurenkov, Yu. A. Mankelevich, and T. V. Rakhimova "Simulation of diffuse, constricted-stratified, and constricted modes of a dc discharge in argon: Hysteresis transition between diffuse and constricted-stratified modes", *Phys. Rev.* E **79**, 046406 (2009).
[8] Y.Z. Ionikh, N.A. Dyatko, A.V. Meshchanov, A.P. Napartovich, F.B. Petrov, "Partial constriction in a glow discharge in argon with nitrogen admixture", *Plasma Sources Sci. Technol*. **21** 055008 (2012)
[9] E. P. Velikhov, V. S. Golubev, and S. V. Pashkin, "Glow-discharge in a gas-flow" *Sov. Phys. Usp*. **25**(5), 340 (1982)
[10] Yu. P. Raizer, *Gas Discharge Physics* (Springer, Berlin, 1991).
[11] Yu. S. Akishev, S. V. Pashkin, and N. A. Sokolov, "Dynamic contraction of steady-state glow discharge in air flow "*Fiz. Plazmy* **4**, 858 (1978).
[12] A. S. Kovalev, I. G. Persiantsev, B. M. Polushkin, A. T. Rakhimov, N. V. Suetin, and M. A. Timofeev,"On the mechanism of the development of breakdown in non-self-sustained gas discharge"*Pis'ma JTP* **6**, 743 (1980).
[13] I. G. Persiantsev, A. T. Rakhimov, N. V. Suetin, and M. A. Timofeev, "Experimental study of the breakdown mechanism in non-self-sustained gas discharge"*Fiz.Plazmy* **9**, 637 (1983).
[14] S. A. Genkin, Y. D. Korolev, G. A. Mesyats, V. G. Rabotkin, and A. P. Khuzeev, "Constriction of a non-self-sustaining spatial discharge ignited by an electron-beam", *High Temp.* **20**, 1 (1982).
[15] V. V. Osipov, "Self-sustained space discharge ", *Phys. Usp.* **43**, 221 (2000).
[16] Yu. S. Akishev, A. M. Volchek, A. P. Napartovich, N. A. Sokolov, and N.I. Trushkin, "On the simulation of the non-uniform contraction in a highly non-linear self-sustained glow discharge" *Fiz. Plazmy* **16**, 474 (1990).
[17] Yu. S. Akishev, A. P. Napartovich, S. V. Pashkin, and N. I. Trushkin, "The measurement of current column plasma parameters in the glow-discharge at increased pressure", *Beitr. Plasmaphys*. **24**, 135 (1984).
[18] Yu. D. Korolev and G. A. Mesyats, *Physics of Pulsed Breakdown in Gases* (URO, Ekaterinburg, Russia, 1998).
[19] G. G. Gladush and A. A. Samokhin, Preprint IAE-3406/6, M., 1981.
[20] B. V. Zhuravlev, A. P. Napartovich, A. F. Pal', V. V. Pichugin, A. V. Rodin, A. N. Starostin, T. V. Taran, M. D. Taran, and A. V. Filippov, "On the nature of the contraction of non-self-sustained gas discharge" *Fiz.Plazmy* **14**, 233 (1988).
[21] K. N. Ulyanov and V. V. Chulkov, "Current-channel formation dynamics in a non-self-maintained discharge plasma", *High Temp.* **21**, 26 (1983).

[22] A. N. Lobanov, V. V. Stepanov, and K. N. Ulyanov, "Dynamics of the current channel in a non-self-sustained discharge plasma", *High Temp*. **22**, 359 (1984).

[23] M.N. Shneider, M.N. Mokrov, G. Milikh, Dynamic Contraction of the Positive Column of a Glow Discharge in Molecular Gas, AIAA-2012-796, 50th AIAA Aerospace Sciences Meeting, Nashville, TN, Jan. 9-12, 2012

[24] M. N. Shneider, M. S. Mokrov, and G. M. Milikh, Dynamic contraction of the positive column of a self-sustained glow discharge in molecular gas, *Physics of Plasmas* **19**, 033512 (2012)

[25] M.N. Shneider, M.N. Mokrov, G. Milikh, Dynamic Contraction of the Positive Column of a Self-Sustained Glow Discharge in Nitrogen/Air Flow, AIAA 2013-1186, 51st AIAA Aerospace Sciences 07 - 10 Jan. 2013, Grapevine, Texas

[26] Yu. P. Raizer, M. N. Shneider, and N. A. Yatsenko, *Radio-Frequency Capacitive Discharges* (CRC, Boca Raton, 1995)

[27] See Compilation of atomic and molecular data, assembled and evaluated by A. V. Phelps: ftp://jila.colorado.edu/collision data/eletrans.txt

[28] I.A. Kossyi, A.Yu. Kostinsky, A.A. Matveyev and V.P. Silakov, "Kinetic scheme of the non-equilibrium discharge in nitrogen-oxygen mixtures", *Plasma Sources Sci. Technol*. **1** 207-220 (1992)

[29] S.O. Macheret, M.N. Shneider, R.B. Miles, R.J.Lipinski, "Electron-beam-generated plasmas in hypersonic magnetohydrodynamic channels", *AIAA J*., **39**, 1127-1138 (2001)

[30] S.O.Macheret, M.N.Shneider, R.B.Miles, "Modeling of air plasma generation by repetitive high-voltage nanosecond pulses", *IEEE Trans. Plasma Sci*. **30**, 1301-1314 (2002)

[31] E. E. Son, "Effect of vibrational temperature on the rate of electronic excitation of diatomic-molecules", *High Temp*. **16**, 990 (1978)

[32] N.L.Aleksandrov, F.I.Vysikailo, R.S.Islamov, I.V.Kochetov, A.P.Napartovich, V.G.Pevgov, "Electron-distribution function in 4:1 N2-O2 mixture", *High Temperature*, **19**(1), 17-21 **(**1981)